\documentclass[conference]{IEEEtran}

\usepackage{cite}
\usepackage{amsmath,amssymb,amsfonts}
\usepackage{algorithmic}
\usepackage{graphicx}
\usepackage{textcomp}
\usepackage{xcolor}
\usepackage{multirow}

\DeclareMathOperator*{\argmin}{arg\,min}
\def\BibTeX{{\rm B\kern-.05em{\sc i\kern-.025em b}\kern-.08em
    T\kern-.1667em\lower.7ex\hbox{E}\kern-.125emX}}

\begin{document}

\title{Long-Term Hail Risk Assessment with Deep Neural Networks}

\author{
\IEEEauthorblockN{Ivan Lukyanenko}
\IEEEauthorblockA{\textit{Department of Control and Applied Mathematics} \\
\textit{Moscow Institute of Physics and Technologies}\\
Moscow, Russia \\
lukianenko.ia@phystech.edu}
\and
\IEEEauthorblockN{Mikhail Mozikov}
\IEEEauthorblockA{\textit{AI Research Center} \\
\textit{Skolkovo Institute of Science and Technology}\\
Moscow, Russia \\
mozikov.mb@phystech.edu}
\and
\IEEEauthorblockN{Yury Maximov}
\IEEEauthorblockA{\textit{Theoretical Division} \\
\textit{Los Alamos National Laboratory}\\
Los Alamos, New Mexico \\
yury@lanl.gov}
\and
\IEEEauthorblockN{Ilya Makarov}
\IEEEauthorblockA{\textit{Industrial AI} \\
\textit{Artificial Intelligence
Research Institute}\\
Moscow, Russia \\
makarov@airi.net}
}
\maketitle

\begin{abstract}

Hail risk assessment is necessary to estimate and reduce damage to crops, orchards, and infrastructure. Also, it helps to estimate and reduce consequent losses for businesses and, particularly, insurance companies. But hail forecasting is challenging. Data used for designing models for this purpose are tree-dimensional geospatial time series. Hail is a very local event with respect to the resolution of available datasets. Also, hail events are rare - only 1\% of targets in observations are marked as "hail".

Models for nowcasting and short-term hail forecasts are improving. Introducing machine learning models to the meteorology field is not new. There are also various climate models reflecting possible scenarios of climate change in the future. But there are no machine learning models for data-driven forecasting of changes in hail frequency for a given area.

The first possible approach for the latter task is to ignore spatial and temporal structure and develop a model capable of classifying a given vertical profile of meteorological variables as favorable to hail formation or not. Although such an approach certainly neglects important information, it is very light weighted and easily scalable because it treats observations as independent from each other. The more advanced approach is to design a neural network capable to process geospatial data. Our idea here is to combine convolutional layers responsible for the processing of spatial data with recurrent neural network blocks capable to work with temporal structure.

This study compares two approaches and introduces a model suitable for the task of forecasting changes in hail frequency for ongoing decades.
\end{abstract}


\section{Introduction}

According to Verisk’s 2021 report \cite{haillosses},  year losses due to hail in 2020 reached \$14.2 billion in the USA. Insurance companies are especially vulnerable to hail events. Urban sprawl and population growth in large cities such as Dallas/Fort Worth, Texas; St. Louis, Missouri; Chicago, Illinois; and Denver, Colorado, have made large amounts of property damage from hail events more likely.

According to \cite{hailform}, hailstones are formed when raindrops are carried upward by thunderstorm updrafts into extremely cold areas of the atmosphere and freeze. Hailstones then grow by colliding with liquid water drops that freeze onto the hailstone’s surface. The hail falls when the thunderstorm's updraft can no longer support the weight of the hailstone, which can occur if the stone becomes large enough or the updraft weakens. In most studies, it is noted that most of the hail's growth occurs at a temperature of approximately -10$^{\circ}$C to -25$^{\circ}$C.

An important ingredient for creating large hailstones is time. Appreciable growth is only attainable if particles remain in an environment conducive to growth for an extended period of time. Some studies suggest that large hailstones spend as much as 10–15 min or more in growth regions of storms.

The hail is an extreme event. The main complexity comes from the fact that hail is very local both in time and space in comparison with the resolution of available climate models. Hail/no hail ratio in the experiment setting is around 1\%. 

For nowcasting (forecasts on a period of up to 2 hours) and short-term (up to 24 hours) hail forecasts, a lot of methods exist.  They are based on various meteorological models and a combination of the latter with machine learning approaches. In \cite{burke2020calibration}, authors add a machine learning model on top of the convection-allowing ensemble system and produce a 24-hour forecast as a result. But models for short-term forecasts differ from models for estimating hail frequencies on a climatological scale. The main reason - is differences in available data. 

Next important in the context of the current research model \cite{PREIN201810} based on a statistical approach to the problem. This model works with joint distribution of atmosphere indices and provides a map of current hail probability distribution around the globe. Although this approach was quite robust and reliable across plain terrains of Europe and the USA, it lacks the ability to take into account landscape influence.  Current research aims to develop a machine learning model capable to work with various climatic models and scenarios of climate change. This model should be able to produce an estimation of changes in hail probability for a given area.   

Our main approach is to use a combination of CNN and LSTM neural networks. It allows the model to catch both spatial and temporal structure of data. 

The baseline approach to compare is to assume an absence of structure due to the sparsity of the target variable and apply gradient boosting.

 The task is to identify favorable to hail development conditions (classification) and then evaluate the frequency of hail events in a given area relying on climate models included in different Coupled Model Intercomparison Project Phases (CMIP5/CMIP6). 

To summarize, our main contributions are as follows:
\begin{itemize}
    \item We  propose a special architecture of a neural network designed to forecast changes in hail frequency for ongoing decades based on CMIP data.
    \item We provide a result of experiments and a comparison with the baseline model. These experiments and comparison show us that our model provides a good approximation of the real annual cycle of hail frequencies (averaged over 2010-2015) and outperforms the baseline approach.  
\end{itemize}

The rest of the work is organized as follows. Section 2 is elaborating on the experiment setting. In this section, we will describe the data we used, quality metrics, and learning settings and formulate a problem statement. In Section 3 we describe the neural network approach to hail forecasting. Here we will introduce our neural network - HailNet. We will show its architecture and explain why we chose this particular type. Section 3 consists of experiment results. In this part, we will create a baseline solution based on gradient boosting and after that compare HailNet performance with the baseline and model introduced in \cite{PREIN201810}.

\section{Related work}

In this section, we will first overview existing approaches to meteorological and climate forecasts. Also, we will overview methods to work with geospatial data in general. 

Machine learning approaches steadily integrate into the meteorological field starting from the operational part \cite{chase2022machine} and reaching narrower domains like agro-meteorology \cite{kelley2020using} or specific weather phenomena like heavy rains \cite{choi2018development}. In agro-meteorology case machine learning is to estimate crop water demand using on-farm sensors and public weather information.

A separate direction is climate studies. The main focus here is on designing or improving global climate models with machine learning techniques involved \cite{dueben2018challenges}, \cite{brenowitz2020machine}. Agro-meteorology is also concerned \cite{cai2019integrating} but on a different scale, here satellite and climate data were integrated to predict wheat yield in Australia using machine learning approaches.

The important idea of why the field of weather forecasts (meteorology) and the field of climate forecasts which seems closely related are yet very different when it comes to machine learning usage is well described in \cite{watson2021machine}.

Let us consider the part of the meteorological field which works with hail prediction. First interesting study here is \cite{burke2020calibration}. The model designed by the authors processes an output of a convection-allowing ensemble system and produces a 24-hour forecast as a result. The possible complication here is that convection-allowing ensembles are not available for almost all countries except for a few. 
The next study is \cite{PREIN201810}. It is based on joint distributions of atmosphere parameters and produces a map of current hail probability distribution around the globe. It is shown in the study that the model has a good generalization ability in Europe and the USA except for mountain areas. 

Our study is related to geospatial time series analysis. Some  approaches to visualize such kind of data are described in \cite{kothur2014interactive}. In the initial steps of research, visualization can help geoscientists to gain a better understanding of geospatial time series. The article \cite{dollner2020geospatial} elaborates on AI-based techniques for 3D point clouds and geospatial data as generic components of geospatial AI.

\section{Experiment setting}
In the first part of this section, we list all sources of data we use and describe preprocessing step. Next, we formulate a problem statement. In the last part, we describe metrics used for evaluating model performance.
\subsection{Data}

Data required to perform hail forecasts arse geospatial time series. Such data are a series of values of a quantity obtained at successive times with equal intervals between them with spatial relations involved. Worth noting here that hailstones formation is associated with cumulonimbus clouds which unlike others have extended vertical structure exceeding in some cases 18 km \cite{Krauss2007}. It is important to address this fact by using data from different isobaric surfaces. So in this way data structure of geospatial time series expands to three dimensions. 

\begin{figure}[htbp]
\centerline{\includegraphics[width=9cm,height=15cm,keepaspectratio]{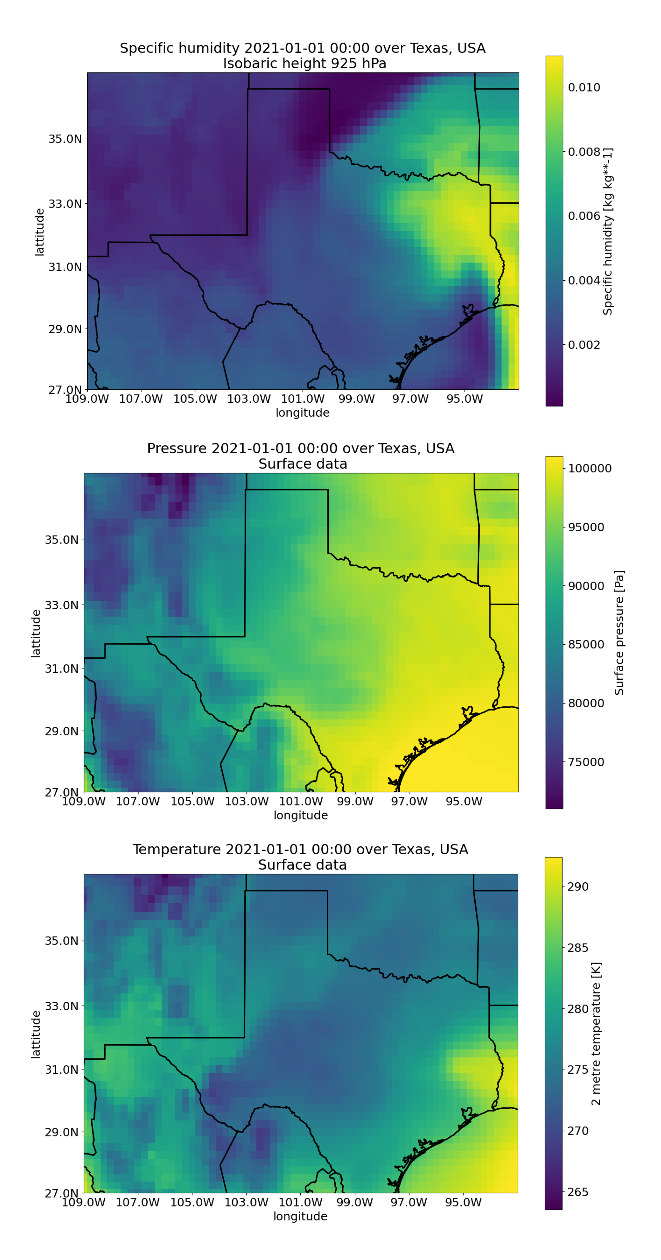}}
\caption{The example of ERA5 data used for training an evaluation of our model. Timestamp: 01.01.2021 00:00. From top to bottom: specific humidity at 925 hPa isobaric surface, pressure at 925 hPa isobaric surface, temperature 2 meters above surface level.}
\label{fig: 3pics}
\end{figure}

The dataset for model training and evaluation includes three sources. The ERA5 hourly data on single levels by European Centre for Medium-Range Weather Forecasts (ECMWF) were used to derive surface climate variables. To address the third dimension of data structure variables from ERA5 hourly data on pressure levels were collected. The train data example is presented in Figure \ref{fig: 3pics}.

The time span is from 2010 to 2021, the time resolution is one hour, model's spatial resolution is $0.25^{\circ}$x$0.25^{\circ}$ latitude and longitude. The target value is binary, either 1: "hail" or 0: "no hail". The National Oceanic and Atmospheric Administration's (NOAA) Storm Events Database was used to collect this data. This database contains information about the occurrence of storms and other significant weather phenomena having sufficient intensity to cause loss of life, injuries, significant property damage, and/or disruption to commerce in the USA. The data concerning the start, end time, and coordinates of hail events were extracted. It is important to note here that on a large timescale the Storm Events Database is not homogeneous due to changes in observation techniques and differences in sources of data. But on the scale of the last decade, this inhomogeneity is almost negligible. Full list of variables is presented in Table \ref{tab: data-list}. 
\begin{table}[htbp]
\caption{Data Description}
\begin{center}
\begin{tabular}{|l|l|c|}
\hline
Dataset & variables & \multicolumn{1}{l|}{Time span} \\ \hline
\multirow{4}{*}{ERA5   on pressure levels} & U-component of wind & \multirow{4}{*}{2010-2021} \\ \cline{2-2}
 & V-component of wind &  \\ \cline{2-2}
 & Specific humidity (calculated) &  \\ \cline{2-2}
 & Temperature &  \\ \hline
\multirow{2}{*}{ERA5 on single levels} & Runoff & \multirow{2}{*}{2010-2021} \\ \cline{2-2}
 & Surface pressure &  \\ \hline
Storm Events   Database & Hail event & \multicolumn{1}{l|}{2010-2021} \\ \hline
\multirow{6}{*}{MRI-CGCM3 RCP8.5} & U-component of wind & \multirow{6}{*}{2022-2050} \\ \cline{2-2}
 & V-component of wind &  \\ \cline{2-2}
 & Specific humidity &  \\ \cline{2-2}
 & Temperature &  \\ \cline{2-2}
 & Runoff &  \\ \cline{2-2}
 & Surface pressure &  \\ \hline
\end{tabular}
\label{tab: data-list}
\end{center}
\end{table}


Coupled Model Intercomparison Project (CMIP) includes climate models introduced by different universities developed under a list of assumptions about climate change scenarios. Scenario Representative Concentration Pathway 8.5 (RCP8.5) is selected. It reflects the modeled climate changes for the case when emissions continue to rise throughout the 21st century. This scenario is a "worst case" option in CMIP5. As follows from the properties of hail events data as dense as possible are needed here.  The suitable dataset is Meteorological Research Institutes' MRI-CGCM3. The spatial resolution of the model is $1^{\circ}$x$1^{\circ}$. The parameters required for the forecast are available in 6 hours time resolution.

Variables describing humidity differ in ERA5 datasets and CMIP data. To coincide with CMIP data, specific humidity was calculated using the MetPy library. Also, there is a mismatch both in spatial and temporal dimensions. It was decided to preprocess all data to the resolutions of the least dense dataset - CMIP5 MRI-CGCM3. The final spatial resolution of all data is $1^{\circ}$x$1^{\circ}$, temporal - 6 hours. 

Also, during preprocessing step it was needed to deal with the problem of imbalanced target data. To address this problem SMOTE technique \cite{chawla2002smote} is applied. After this step ratio of hail/no hail events has increased from 1/100 to 1/1.
\subsection{Problem statement}
Let us give some designations and formalize our problem:

$n \in \mathbb{N}$ -- number of climatic variables;

$\text{long} \in \mathbb{N}$ -- longitude of considered area;

$\text{lat} \in \mathbb{N}$ -- latitude of considered area;

$X_{i,j,k} \in \mathbb{R}^{\text{n}\times \text{lat} \times \text{long}} $ -- a tensor of climatic variables at one timestamp;

$\widetilde{X}_{t, i, j, k}$ -- a time-series of climatic variables corresponding to one day;

$f(w, X)$ -- predicting model.

We are solving the time series classification problem for every point on the grid of the considered region. Our objects are time series of tensors $X_{i,j,k}$ corresponding to every 6-period. Our targets are $\text{lat} \times \text{long} $ matrices with zeros (no hail) and ones (hail) corresponding to hail events. 

The output of the $f(w, x)$ is 2-dimensional probability pseudo-distribution. Every output value range from 0 to 1, but sum of every matrix value is not equal to 1, because favorable conditions for hail formation can occur on the same day in many points at the same day and vice versa.

Mathematically we set the optimization problem of minimizing the Mean Squared Error between output pseudo-distributions and target grids for finding the best parameters for our model.

\begin{equation}
    w^* = \argmin\limits_w\frac{1}{N}\sum\limits_{t = 1 }^{N}\|f(w, X_t) - Y_t\|^2
\end{equation}

\subsection{Quality metrics}
We will use two basic (precision (\ref{eq: pre}) and recall (\ref{eq: rec})) and one domain-specific metric (\ref{eq: rmse}).
\begin{equation}
 Precision = \frac{TP}{TP+FP} 
 \label{eq: pre}
\end{equation}
\begin{equation}
 Recall = \frac{TP}{TP+FN}
 \label{eq: rec}
\end{equation}
Where:

$TP$ - true positive rate of model predictions;

$FP$ - false positive rate of model predictions;

$FN$ - false negative of model predictions.

Talking about the domain-specific metric, it is highly important to ensure that our model is capable to catch the annually-periodic structure of hail occurrence. It is possible with a special metric firstly introduced in \cite{PREIN201810}. The metric is a simple root-mean-squared error ($RMSE_{om}$) between the normalized observed and modeled annual cycle of monthly mean hail frequency over the target area.

\begin{equation}
    RMSE_{om} = \sqrt{\sum_{i=1}^{12}(o_{m}-m_{m})^2}
 \label{eq: rmse}
\end{equation}
Where:

$o_{m}$ - normalized observed annual cycle of monthly mean hail frequency;

$m_{m}$ - normalized modeled annual cycle of monthly mean hail frequency;

Normalization is needed due to the fact that the dataset with hail observation is partially inhomogeneous. Although this inhomogeneity is small during the 2010-2021 years, we still preserve normalization to increase further stability of metric. Also, this approach coincides with the original paper, so we can compare models' performance.

Let us denote classes of target data:
\begin{itemize}
    \item 1: "hail"
    \item 0: "no hail"
\end{itemize}

In the current setting, it makes sense to pay more attention to the recall metric of our model. The cost of false negatives (the case when we underestimated hail frequency) is expected to be higher than of false positives.

\section{Neural network approach (HailNet)}

\begin{figure}[htbp]
\centerline{\includegraphics[width=11cm,height=11cm,keepaspectratio]{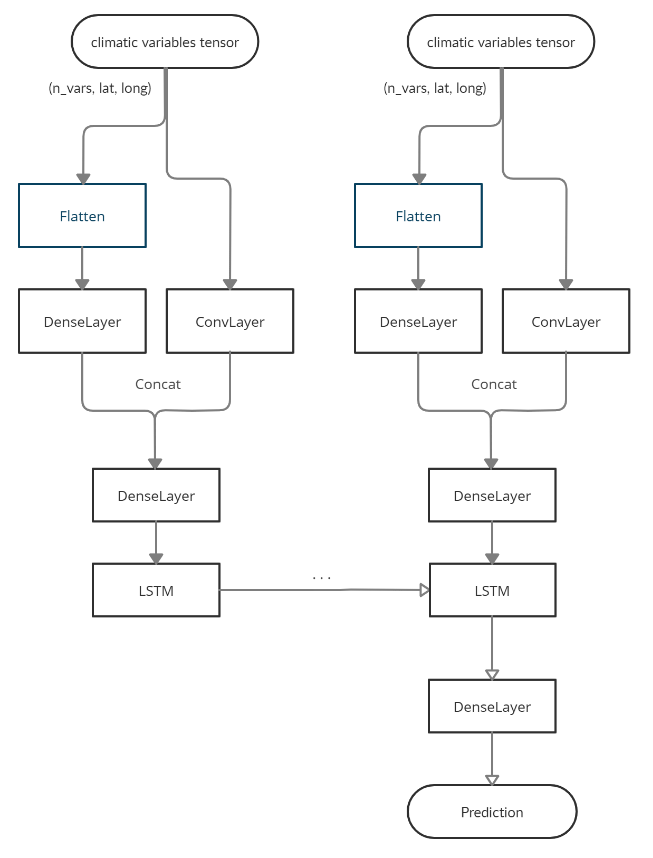}}
\caption{Architecture of HailNet neural network}
\label{fig: HailNet2.png}
\end{figure}

The next step is to build a model capable to catch the spatial and temporal structure of climate variables. It's possible by introducing a neural network with an architecture presented in Figure \ref{fig: HailNet2.png}. Further in the paper, it will be referred to as HailNet.
The neural network has two steps of feature extraction:
\begin{itemize}
\item Convolutional (Conv2d) + dense layers
\item Recurrent layers (LSTM) 
\end{itemize}

The first step allows us to catch the spatial structure of data. After that, the concatenation of outputs from Dense and Convolutional layers maps every climate tensor of time series to embedding space. Objects from embedding space have fewer dimensions and are specified with important spatial information extracted from raw climate tensors. Further, we observe time series of elements from embedding space. Using the LSTM network (second step) HailNet extracts temporal features from time series. The output of the last LSTM unit proceeds to the Dense layers with a sigmoid activation function on the output. As an output, we get a 2-dimensional grid with shape lat $\times$ long with values in a range from 0 to 1. This grid is interpreted as a 2-dimensional probability pseudo-distribution.

\section{Experiment results}

In this section, we describe a baseline approach to the problem of hail forecasting. Later, we compare the performance of the HailNet to the baseline model and to a model introduced in  \cite{PREIN201810}. In the last part, we present a sample of hail frequency forecast based on the CMIP5 MRI-CGCM3 (RCP85) model for Texas, part of New Mexico, and Oklahoma.
\subsection{Baseline approach description}
At the beginning of our experiment, we decided to create a simple baseline solution to refer to. This solution to the problem is to group all features in bounds of one day and assume an absence of temporal and spatial structure in this timescale. The assumption is based on the fact that hail events are too local. It is enough to observe 24 hours to have all features corresponding to a single hail event. 

Worth noting that ignoring spatial structure prevents the model to have any information about associated synoptic processes, precisely, atmospheric fronts. But hail-producing thunderstorms are quite often associated with cold atmospheric fronts. Still, it is not obvious whether data with a resolution of $1^{\circ}$x$1^{\circ}$ still preserve spatial features corresponding to processes above the synoptic scale properly. And baseline assumption here - it is not. To train the baseline model CatBoost classifier \cite{dorogush2018catboost} is used. 

The parameters of the model are the following:
\begin{itemize}
\item iterations=1000
\item learning\_rate=0.01
\end{itemize}

The results of the baseline approach will be compared with HailNet in the next part of this section.

\subsection{HailNet}

The first set of metrics we want to refer to is precision and recall. Values of those metrics both for the baseline CatBoost solution and the HailNet are presented in Table \ref{tab: Contingency table}, but values for the model from \cite{PREIN201810} are unavailable. As was stated before, we aimed to put more value to recall maximization, but still, it is not possible to ignore the precision metric. 

Although we reached high values of metrics in focus for class 0: "No hail" the real interest is in 1: "Hail", because this class was initially underrepresented and was synthetically sampled with SMOTE. We observe good recall values for this class both for the baseline solution and the HailNet. This means that a great portion of real hail events was correctly labeled by our classifiers and the number of false negatives is small. Yet the baseline solution has remarkably small precision. This means that the model tries to be "overcautious" and labels a lot of actual 0: "No hail" events as 1: "Hail". 

The HailNet shows a gain in precision with respect to the baseline model. It reaches 0.2 which is still not a high value of the metric. But in the setting of the experiment, this is acceptable. 

\begin{table}[htbp]
\caption{Precision and recall metrics of HailNet and baseline model }
\begin{center}
\begin{tabular}{|cll|}
\hline
\multicolumn{1}{|c|}{} & \multicolumn{1}{l|}{Precision} & Recall \\ \hline
\multicolumn{3}{|c|}{HailNet} \\ \hline
\multicolumn{1}{|c|}{No hail: 0} & \multicolumn{1}{l|}{0.98} & 0.76 \\ \hline
\multicolumn{1}{|c|}{Hail: 1} & \multicolumn{1}{l|}{0.2} & 0.79 \\ \hline
\multicolumn{3}{|c|}{CatBoost baseline} \\ \hline
\multicolumn{1}{|c|}{No hail: 0} & \multicolumn{1}{l|}{0.99} & 0.68 \\ \hline
\multicolumn{1}{|c|}{Hail: 1} & \multicolumn{1}{l|}{0.02} & 0.77 \\ \hline
\end{tabular}
\label{tab: Contingency table}
\end{center}
\end{table}

\begin{figure}[htbp]
\centerline{\includegraphics[width=9cm,height=9cm,keepaspectratio]{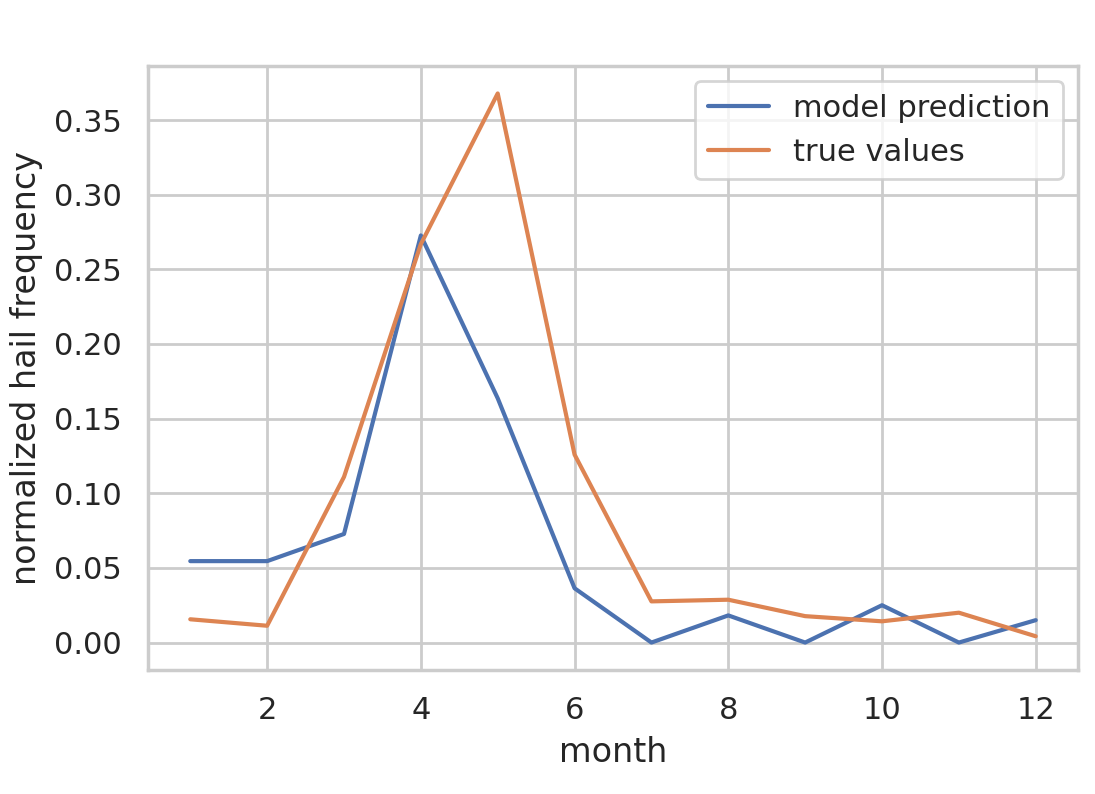}}
\caption{Normalized observed and modeled annual cycle of monthly mean hail frequency}
\label{fig: normalized-annual-cycle}
\end{figure}

To ensure that our model is capable to catch the annually-periodic structure of hail occurrence RMSE between the normalized observed and modeled annual cycle of monthly mean hail frequency over the target area was calculated. The actual shapes of corresponding curves are depicted in Figure \ref{fig: normalized-annual-cycle}. We see that the HailNet approximates all months except June quite well. The June gap is a matter of further research.

The baseline model scored 0.19 for $RMSE_{om}$ which is already close to the values stated in \cite{PREIN201810}. Scores from the reference are approximately 0.18 in the best case and 0.20 for the model which is considered optimal with respect to a number of features involved. The HailNet achieved 0.16 and slightly outperformed \cite{PREIN201810} model in the $RMSE_{om}$ metric.

\subsection{CMIP-based forecast}

\begin{figure}[htbp]
\centerline{\includegraphics[width=9cm,height=9cm,keepaspectratio]{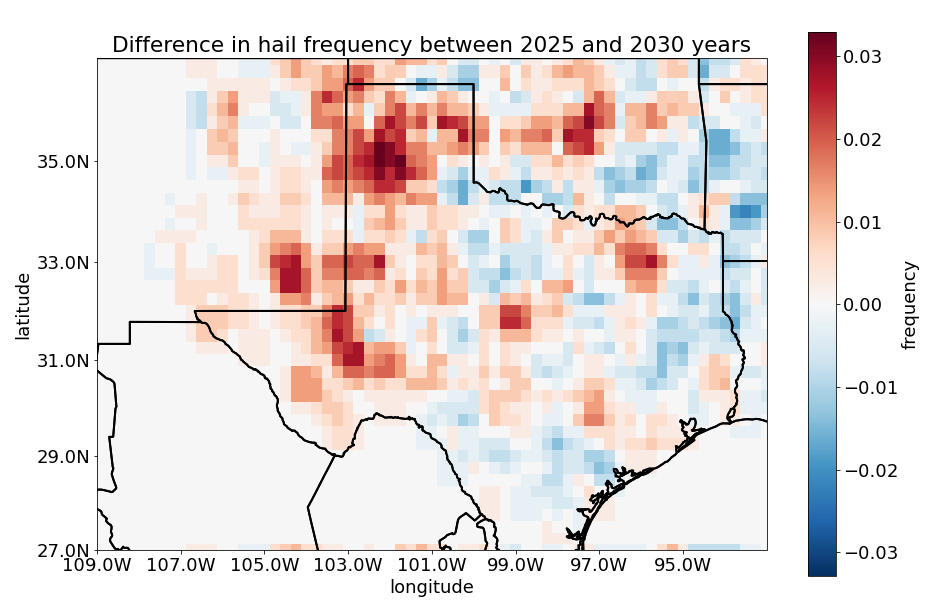}}
\caption{Difference in hail frequency between 2030 and 2025 years}
\label{fig: delta25-30}
\end{figure}

Using data from CMIP5 MRI-CGCM3 model (scenario RCP8.5) we created a forecast of hail frequency. The sample of it for Texas, part of New Mexico, and Oklahoma for 2025 and 2030 years is presented in Figure \ref{fig: delta25-30}. The hail frequency heat maps exhibit a clear persistent peak in Northern Texas. This  peak is probably associated with the leeward side of mountains located to the West. The extent and direction of this influence are a matter of further research. 
\begin{figure}[htbp]
\centerline{\includegraphics[width=9cm,height=9cm,keepaspectratio]{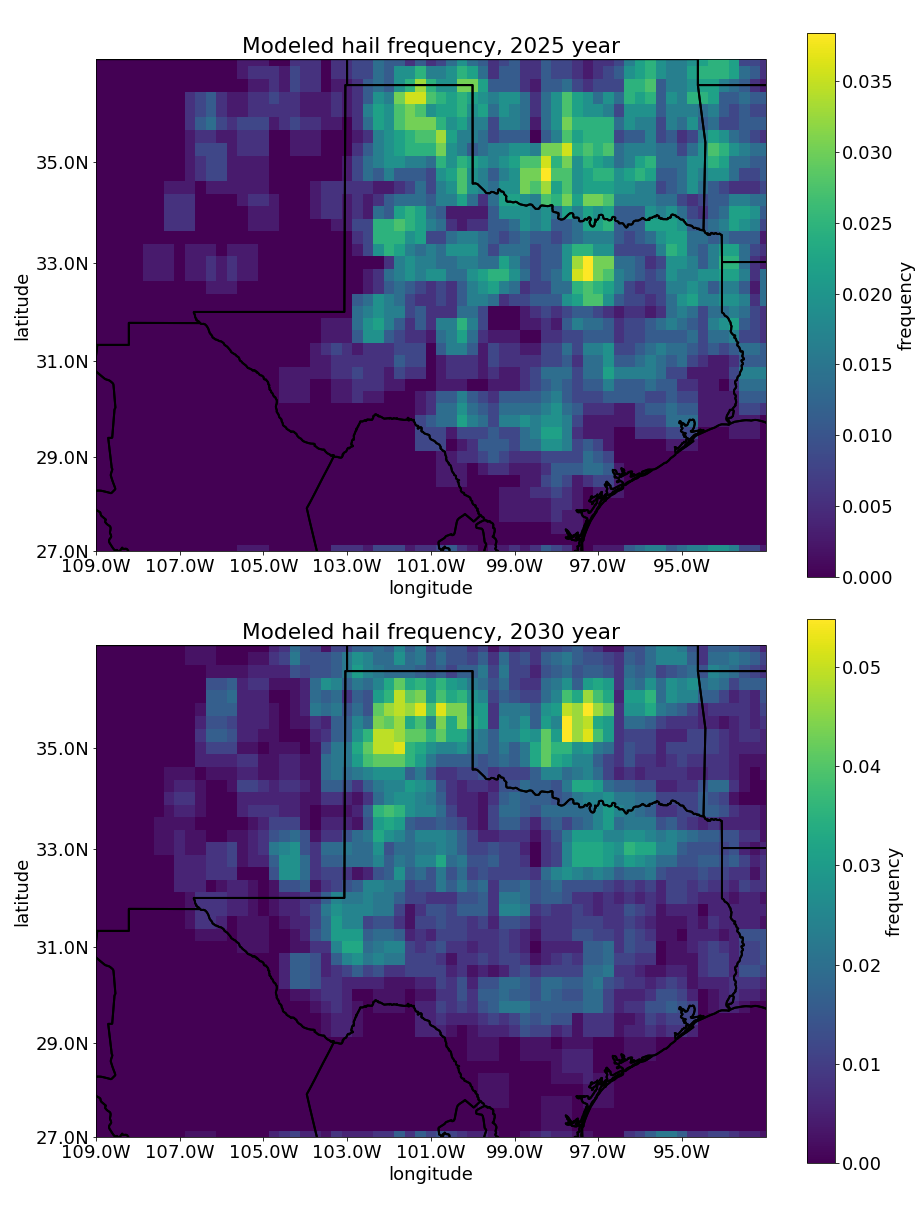}}
\caption{Modeled hail frequency for 2025 and 2030 years based on MRI-CGCM3 (RCP8.5)}
\label{fig: modeled25-30}
\end{figure}

In the plain area, the pattern is more complicated. The hail frequency at some points may be comparable to the Northern counterpart, but there is no persistent geographical location for it.

In Figure \ref{fig: delta25-30} which depicts the difference in hail frequency between 2030 and 2025 years, we observe an increasing trend  in the leeward area 
and high variability in frequency difference in plain area.

\section{Conclusion}

In this paper, we introduced a new model - HailNet, designed to produce climatological hail forecasts based on CMIP data. It slightly outperformed the model, designed for a close purpose \cite{PREIN201810}.

In the final part of the paper, we presented a sample of hail forecast for Texas, part of New Mexico, and Oklahoma and performed a brief analysis of it.

Hail forecasting is a challenging task. One of the main reasons is the sparsity of the target. And the complexity even grows when scientists do not have a proper dataset of hail observations. It is the case for almost all countries except the USA and some parts of the European Union. But even with those available datasets, one should work with caution for a number of reasons. Single dataset (as with Storm Events Database) may be compiled from different sources. This fact introduces inhomogeneity to data which in particular cases (e.g. fast expansion of weather station coverage) may be quite severe. Observation techniques and quality of observations, in general, improve through time. It creates an artificial increasing trend. All these factors should be treated while working with historical data.

But with side effects aside, improvements in meteorological stations' equipment, and expansion of station coverage will certainly yield a great benefit to hail forecasting in particular and meteorological forecasts in general.

\bibliographystyle{plain}
\bibliography{refs.bib}

\end{document}